\documentclass[10pt]{IEEEtran}

\ifCLASSINFOpdf

\fi

\usepackage[subrefformat=parens,labelformat=parens,caption=false,font=footnotesize]{subfig}
\usepackage{array}
\usepackage{balance}
\usepackage{amsmath}
\usepackage{breqn}
\usepackage[nolist,nohyperlinks]{acronym}
\usepackage{graphicx,wrapfig,lipsum}
\usepackage{rotating}
\usepackage{amsmath, amssymb, amsthm}
\usepackage{stmaryrd}
\usepackage{tikz}
\usepackage{verbatim}
\usepackage{url}
\usepackage[utf8]{inputenc}
\usepackage[english]{babel}

\usepackage{multicol}
\usepackage{xr-hyper}

\usepackage{scalerel}
\usepackage{multicol}

\usepackage[noadjust]{cite}
\usepackage[normalem]{ulem}
\usepackage{color}
\usepackage{dsfont}
\usepackage{bm}
\usepackage[short]{optidef} 
\usepackage{diagbox}
\usepackage{enumitem}
\usepackage{slashbox}
\usepackage[ruled]{algorithm2e}
\usepackage{multirow}
\usepackage[T1]{}
\usepackage{color, colortbl}
\usepackage{babel}
\usepackage{verbatim}
\usepackage{textcomp}
\usepackage{tabulary}
\usepackage[pscoord]{eso-pic}
\newcommand{\placetextbox}[3]{
  \setbox0=\hbox{#3}
  \AddToShipoutPictureFG*{
    \put(\LenToUnit{#1\paperwidth},\LenToUnit{#2\paperheight}){\vtop{{\null}\makebox[0pt][c]{#3}}}%
  }%
}%

\usepackage{booktabs}
\usepackage{caption}
\definecolor{Gray}{gray}{0.95}
\definecolor{LightCyan}{rgb}{0.8,0.85,1}
\definecolor{LightBlue}{rgb}{0.6,0.6,1}
\usepackage[font=small]{caption}
\usepackage{mathtools}

\setlist{nosep}

\usepackage{mathtools}

\begin{document}

\title{Machine Learning and Analytical Power Consumption Models for 5G Base Stations}

\author{Nicola Piovesan, David L\'opez-P\'erez, Antonio De Domenico, Xinli Geng,\\  Harvey Bao, and M\'erouane Debbah\\
 \small{Huawei Technologies, Paris Research Center, 20 quai du Point du Jour, Boulogne Billancourt, France.}
 \\ \small{Email: nicola.piovesan@huawei.com}}
\maketitle
\thispagestyle{empty}

\placetextbox{0.5}{0.05}{\textsf{\scriptsize © 2022 IEEE. Personal use of this material is permitted. Permission from IEEE must be obtained for all other uses, in any current or future media, including }}
\placetextbox{0.5}{0.04}{\textsf{\scriptsize  reprinting/republishing this material for advertising or promotional purposes, creating new collective works, for resale or redistribution to servers or lists,}}%
\placetextbox{0.5}{0.03}{\textsf{\scriptsize  or reuse of any copyrighted component of this work in other works.}}%
\placetextbox{0.5}{0.98}{\textsf{\scriptsize N. Piovesan, D. David L\'opez-P\'erez, A. De Domenico, X. Geng, H. Bao, M. Debbah, "Machine Learning and Analytical Power Consumption Models for 5G Base Stations",}}
\placetextbox{0.5}{0.97}{\textsf{\scriptsize IEEE Communications Magazine, October 2022. DOI: 10.1109/MCOM.001.2200023}}

\begin{abstract}
The energy consumption of the fifth generation (5G) of mobile networks is one of the major concerns of the telecom industry.
However, there is not currently an accurate and tractable approach to evaluate 5G  
base stations (BSs) power consumption. 
In this article, 
we propose a novel model for a realistic characterisation of the power consumption of 5G multi-carrier BSs, 
which builds on a large data collection campaign. 
At first, 
we define 
a machine learning architecture that allows modelling multiple 5G BS products. 
Then, 
we exploit the knowledge gathered by this framework to derive a realistic and analytically tractable power consumption model,
which can help driving both theoretical analyses as well as feature standardisation, development and optimisation frameworks. 
Notably, we demonstrate that such model has high precision, 
and it is able of capturing the benefits of energy saving mechanisms. 
We believe this analytical model represents a fundamental tool for understanding 5G BSs power consumption, 
and accurately optimising the network energy efficiency.
\end{abstract}

\section{Introduction}
\label{sec:intro}

The \ac{5G} of radio technology has brought about new services, technologies, and networking paradigms,
with the corresponding societal benefits.
However, the energy consumption of the new \ac{5G} network deployments is concerning.
Deployed 5G networks have been estimated to be about 4$\times$ more energy efficient than 4G ones.
Nonetheless, their energy consumption is around 3$\times$ larger,
due to the larger number of cells needed to provide the same coverage at higher frequencies,
and the increased processing required by its wider bandwidths and more antennas~\cite{Huawei2020}.
To put this number into context,
it should be noted that, in average, the network \ac{OPEX} already accounts for around 25\,\% of the total operator’s cost, and that 90\,\% of it is spent on large energy bills~\cite{GSMA20205Genergy}.
Notably,
most of this energy ---more than 70\,\%--- has been estimated to be consumed by the \ac{RAN},
and in more detail, 
by the \acp{BS},
while data centres and fibre transport only account for a smaller share~\cite{GSMA2021,lopezperez2022survey}.
	
To decrease the \ac{RAN} energy consumption,
\ac{3GPP} \ac{NR} Release 15 specified intra-\ac{NR} network energy saving solutions, 
similar to those developed for \ac{3GPP} \ac{LTE},
e.g., autonomous cell switch-off/re-activation capabilities for capacity booster cells via X$_n$/X$_2$ interfaces. 
Moreover, \ac{3GPP} \ac{NR} Release 17 has recently specified inter-system network energy saving solutions,
and is currently taking network energy saving as an artificial intelligence use case.
However, data gathered about the benefits brought by \ac{3GPP} \ac{LTE} and \ac{NR} energy saving solutions have shown that they are not enough to fundamentally address the energy consumption challenge~\cite{CT2021report}. 

To continue tackling this challenge, 
\ac{3GPP} \ac{NR} Release 18 has recently approved a new study item, titled \emph{``Study on \ac{NR} network energy saving enhancements''}, 
which attempts to develop a set of more flexible and dynamic network energy saving solutions~\cite{CT2021report}.
In more details, 
the main objectives of this study item are:
\begin{enumerate}
    \item
    Identify new energy saving scenarios beyond that of the capacity booster cell, 
    e.g., compensation cells;
    \item
    Study enhancements to allow a faster adaptation of networking resources to traffic needs through, 
    e.g., \emph{i)} \ac{UE} assistance information reports, 
    \emph{ii)} \ac{BS} information exchange to share traffic predictions and support both beam-level operation and transmit power adjustment coordination, and  
    \emph{iii)} \ac{DL}/\ac{UL} channel measurement enhancements.
\end{enumerate}

Importantly, 
to analyse the gains brought by such new schemes, 
there has been consensus on the need for new models to accurately estimate the \ac{5G} network power consumption. 
\ac{3GPP} \ac{NR} Release 16 defined a power consumption model for \ac{5G} \acp{UE} \cite{Kim2020}. 
However, there is no \ac{5G} network counterpart. 
Ongoing \ac{3GPP} discussions have suggested that such new \ac{5G} network power consumption model should be a function of the number of \acp{BS} in the area of study, their frequency of operation, bandwidth, transmit power, number of transceivers, signalling configuration, \acp{PRB} load, \ac{MIMO} layers usage, as well as energy saving functionalities and their related sleep states and transition times.

To fill this gap, 
in this paper, 
we introduce a new power consumption model for \ac{5G} \acp{AAU}, the highest power consuming component of a \ac{BS}\footnote{
In \ac{5G} terminology, 
a massive \ac{MIMO} \ac{BS} is divided into three parts: the centralised unit, the distributed unit and the \ac{AAU}.}
and in turn of a mobile network. 
In particular, 
we present 
an analytically tractable model,
which builds on a large data collection campaign and our \ac{ML}-based analysis. 
The proposed model is realistic, 
as it is characterised by a high precision, 
and generalises well to a high number of \ac{5G} \ac{AAU} types/products.
For example, 
it accounts for multi-carrier \acp{AAU} embracing the widely used \ac{MCPA} technology~\cite{zhang2019mcpa}.\footnote{ 
An \ac{MCPA} operates, 
in contrast to a single-carrier \ac{PA}, 
on multiple carriers as input, 
and provides a single amplified output.} 
This allows to share some of the \ac{PA} hardware among multiple carriers managed by an \acp{AAU}, 
thus reducing its power consumption.
Moreover, 
our model also captures the benefits brought by complex, standardised shutdown schemes, 
i.e., carrier shutdown, channel shutdown, symbol shutdown, and deep dormancy~\cite{lopezperez2022survey}, 
when operating in the field. 

About the methodology adopted in this paper, 
it should be highlighted that the parameters of the proposed analytical model are derived for a selected AAU product by using data collected from a real network deployment. 
Unfortunately, however, it is generally not possible to obtain exhaustive data for all possible input configurations for all AAU products deployed in real networks. 
Importantly, the inaccessibility of AAU measurements of power consumption under some conditions may prevent the derivation of the analytical model parameters.
Therefore, we implemented a methodology, 
in which a \ac{ML} framework is designed and trained to gather knowledge from many different types of AAU with different hardware configurations.
Notably, this modeling approach allows,
taking advantage of the \ac{ML} generalization properties,
generating synthetic data covering scenarios that may not be directly observable in the collected data but that are needed to derive the proposed analytical model.

\section{Related Works on \ac{BS} Power Model}
\label{sec:sota}

It has been reported that 73\,\% of the total network energy is consumed by the \acp{BS}~\cite{GSMA2021},
where the power amplifier, the transceivers and the cables consume about 65\,\% of the total \ac{BS} energy~\cite{lopezperez2022survey}. 
Therefore, 
significant attention has been directed towards reducing the energy consumed by the \acp{BS} during the last years, 
and various \ac{BS} power consumption models have been proposed and investigated,
as a result. 

The work in~\cite{Auer2011} proposed one of the most widely used \ac{BS} power consumption models in the literature. 
In particular, 
such model explicitly shows the linear relationship between the \ac{BS} power consumption and its transmit power.
Embracing the model in~\cite{Auer2011},
the work in \cite{Debaillie2015} proposed an extension, 
which additionally supports \acp{mMIMO} and energy saving capabilities, 
considering different sleep depths and transition times between different energy states.
However, multi-carrier and/or \ac{CA} capabilities were not considered,
and \acp{mMIMO} power consumption estimations seem inaccurate~\cite{han2020energy},
with an optimistic 40.5\,W per \ac{BS}.

With regard to \ac{mMIMO}, 
the work in~\cite{Tombaz2015} extended the \ac{BS} power consumption model in~\cite{Auer2011},
considering a linear increase of the power consumption with the number of \ac{mMIMO} transceivers.
More advanced works followed in this area,
highlighting the importance of taking the impact of multi-\ac{UE} scheduling and other \ac{mMIMO} \ac{BS} components into account in the modelling of \ac{5G} \ac{BS} power consumption, 
such as power amplifiers, transceivers, analog filter and oscillators. 
Specifically, 
the cornerstone research in~\cite{bjornson2015optimal} provided a more complete model, 
which considers the \ac{mMIMO} \acp{BS} architecture, 
both \ac{DL} and \ac{UL} communications, 
as well as the number of \acp{UE} multiplexed per \ac{PRB}, 
and a large number of \ac{mMIMO} \ac{BS} components. 

When modelling the power consumption of a system using multiple carriers and/or \ac{CA}, 
it is also necessary to take into account how the power consumption scales with the number of \acp{CC} managed by the \ac{BS}.
The work in~\cite{Yu2015} captures this relationship using a linear model,
but the literature is sparse in this area.

The work in~\cite{lopez2021energy} further combined and extended the linear version of the above presented works,
jointly considering \ac{mMIMO} and multi-carrier capabilities features, 
such as \ac{CA} and its different aggregation capabilities, 
i.e., intra-band contiguous, intra-band non-contiguous and inter-band. 

\section{5G AAU architecture model}
\label{sec:system_model}

Although various aspects related to the power consumption of a \ac{5G} \ac{BS} have been considered in the research presented in Sec.~\ref{sec:sota}, 
the true complexity of a 5G multi-carrier \ac{mMIMO} \ac{AAU},
where a single power amplifier may accommodate for multiple carriers, 
using \ac{MCPA} technology, 
is not embraced in any of them.
Our paper fills this gap,
by defining a general and practical \ac{AAU} architecture,
and providing first the corresponding data-driven power consumption model,
and then an analytical formulation fitted with realistic values for a particular \ac{AAU} type. 

\begin{figure}[t!]
    \centering
    \includegraphics[width=8.4cm]{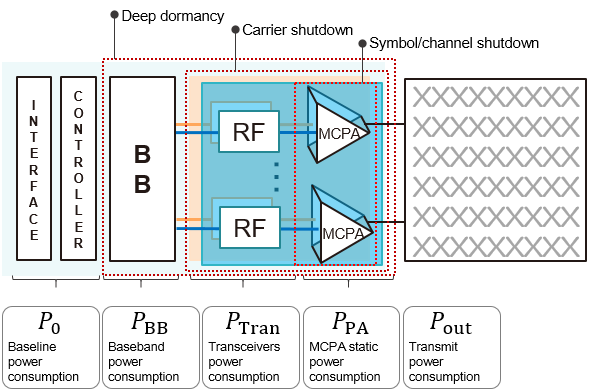}
    \caption{AAU with \acp{MCPA} handling 2 \acp{CC} in 2 different bands,
    which transmits over the same wideband antenna panel. 
    }
    \label{fig:aau}
\end{figure}

In more detail, 
in our \ac{AAU} architecture, 
we assume that:
\begin{itemize}
\item The AAU has a multi-carrier structure, 
and uses \ac{MCPA} technology;
\item The \ac{AAU} manages $C$ carriers ---or \acp{CC}, using a \ac{CA} terminology--- deployed in $T$ different frequency bands;
\item The \ac{AAU} comprises $T$ transceivers, 
each operating a different frequency band, 
and $M$ MCPAs, one for each antenna port;
\item A 
transceiver includes $M$ \ac{RF} chains, 
one per antenna port, 
which comprehend a cascade of hardware components for analog signal processing, 
such as filters and digital-to-analog converters;
\item Antenna elements are assumed to be passive. 
For example, one wideband panel or $T$ antenna panels may be used per \ac{AAU};
\item Deep dormancy, carrier shutdown, channel shutdown, and symbol shutdown are implemented, 
each switching off distinct components of the \ac{AAU}. 
\end{itemize}

Fig. 1 shows the AAU architecture and its main power consumption components. 
In more details, 
the overall AAU consumed power includes: 
\emph{i)} 
the baseline power consumption, $P_0$,
which accounts for part of the AAU circuitry that is always active 
(e.g., circuitry used to control the AAU activation/deactivation),
\emph{ii)} 
the power consumption, $P_{\rm BB}$, required for the baseband processing performed at the \ac{AAU},
\emph{iii)} 
the power consumed by the $T$ transceivers in the AAU,
i.e., $P_{\mathrm{Tran}}$,
\emph{iv)} 
the static power consumed by the \acp{MCPA}, 
i.e., $P_{\mathrm{PA}}$, and
\emph{v)} 
the power consumed to generate the transmit power required to transmit the data over the $C$ CCs, 
i.e., $P_{\mathrm{out}}$.

As described in~\cite{zhang2019mcpa}, 
it should be highlighted that the implementation of \acp{MCPA} results in an increased energy efficiency with respect to single-carrier power amplifiers. 
In more details, 
by integrating multiple carriers together, 
the total transmit power managed becomes greater, 
thus enabling \acp{MCPA} to operate at higher efficiency areas.
Moreover, the static power consumption of the \acp{MCPA} increases sub-linearly with respect to the number of carriers,
as part of the signal processing components can be shared among them. 
However, it is worth highlighting that the implementation of \acp{MCPA} entails increased complexity in the management of the network energy saving, 
and thus, 
in the estimation of the power consumption. 
In fact, contrarily to what commonly considered by simplistic models, 
the deactivation of just one carrier may not bring the expected energy savings, 
if the \acp{MCPA} need to remain active to operate other co-deployed active carriers.

\section{Artificial Neural Network model}
\label{sec:ml}

In this section, 
we describe the measurements gathered during our data collection campaign. 
Moreover, 
we provide a detailed description of the implemented \ac{ANN} architecture for modelling and estimating power consumption, 
as well as an analysis of its accuracy. 
Note that ANNs were selected after evaluating and comparing their performance with those of other ML methods. 
The better performance of ANNs emanates due to their better capabilities to deal with the available tabular data and superior generalisation properties. 

\vspace{-0.15cm}
\subsection{Dataset}
We collected hourly measurements for 12 days from a real deployment with 7760 \ac{5G} \acp{AAU} in China, comprising 25 different types of \ac{AAU} from a single vendor.
Note that such data contains sensitive information regarding proprietary product hardware specifications, which cannot be made publicly available.
The gathered information contains 150 different features,
which can be divided into four main categories:
\begin{itemize}
    \item \textit{Engineering parameters}: 
    Information related to the configuration of each \ac{AAU} 
    (e.g., \ac{AAU} type, number of \ac{RF} chains, numbers of supported and configured carriers);
    \item \textit{Traffic statistics}: 
    Information on the serviced traffic 
    (e.g., average number of active \acp{UE} per transmission  time interval, number of used \acp{PRB}, traffic volume);
    \item \textit{Energy saving statistics}: 
    Information on the activated energy saving modes 
    (e.g., duration of the carrier, channel and symbol shutdown as well as dormancy activation);
    \item \textit{Power consumption statistics}: 
    Information on the power consumed by the \acp{AAU}.
\end{itemize}

\vspace{-0.15cm}
\subsection{Inputs of the model}

Feature importance analysis was performed to identify the most relevant input features in the available dataset. 
Such features are the type/model of \ac{AAU}, 
together with the key characteristics of the configured carriers.
To give an example,
such key characteristics comprehend,
among others,
frequency- and power-related engineering parameters, 
such as the carrier frequency, bandwidth and transmit power, 
the \ac{DL} \ac{PRB} load, 
and the amount of time for which each energy saving mode is activated. 
See Fig.~\ref{fig:MLmodel} for a detailed description of all the selected input features. 
Note that the identified features are fundamental parameters,
which are available in the products of any vendor. Moreover, feature importance analysis can extend the inputs of our ANN model to consider proprietary and not standardised energy saving schemes.

After selection,
each of the input features was pre-processed, 
and then represented by one or more neurons at the input layer of the \ac{ANN}.
The numerical features were normalised before being input to the model, 
whereas the categorical ones were input by using one-hot-encoding.

Since a \ac{5G} \ac{AAU} can operate multiple carriers trough a \ac{MCPA},
to make our \ac{ANN} model the most general and flexible,
the input layer takes input from the maximum number of carriers that can be managed by the most capable \ac{AAU} in the dataset.
When less carriers are deployed in an \ac{AAU}, 
the input neurons related to the none deployed carriers are set to zero.
This approach allows to implement our \ac{ANN} model with a fixed number of input neurons, 
and thus construct a single model for all possible \ac{AAU} types and carrier configurations,
with a minimal accuracy loss. 
The maximum accuracy loss observed when comparing this single model approach with respect to training different models for the different \ac{AAU} types and carrier configurations is 1.86\,\%.

\vspace{-0.2cm}
\subsection{Outputs of the model}
\label{sec:output}

Different power consumption values are observed in the data for the same input feature values due to missing input features and/or errors in the measurements or in the collection/processing of the data.
To embrace such noise,
we define the measured power consumption 
in our \ac{ANN} model as the expected power consumption for a given input configuration plus a noise,
originating from the mentioned errors.
The analysis of the available data highlighted that such noise is normally distributed. 
It thus follows that the measured power consumption is normally distributed. 

With this in mind, the goal of our \ac{ANN} model is to produce, for a given input configuration, an estimate of the mean and standard deviation of the power consumption distribution.
This allows having an evaluation of the confidence interval for each of the performed power consumption estimations during training and testing,  
and in turn, 
increase the reliability of the whole estimation process. 

\vspace{-0.15cm}
\subsection{Model architecture}

\begin{figure}[t!]
    \centering
    \includegraphics[width=8.5cm]{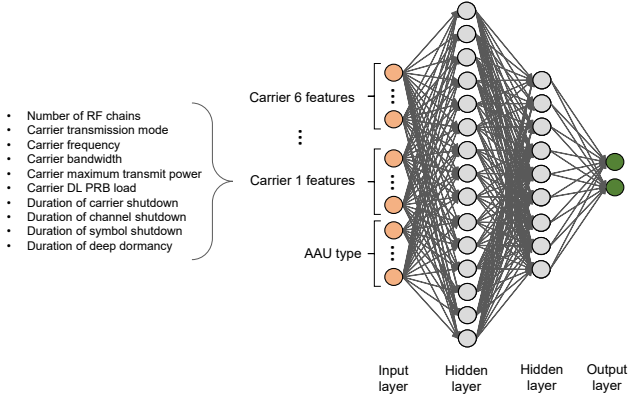}
    \caption{Architecture of the designed \ac{ANN}.}
    \label{fig:MLmodel}
\end{figure}

We consider the multilayer perceptron as basic architecture for the \ac{ANN}, 
consisting of multiple fully connected layers of neurons~\cite{Goodfellow-et-al-2016}. 
The overall architecture of the proposed \ac{ANN} model is also depicted in Fig.~\ref{fig:MLmodel}.

In our specific scenario, 
we collected data for 25 different \ac{AAU} types, 
where the most capable \ac{AAU} supports up to 6 \acp{CC}. 
The input layer is thus composed of 85 neurons, 
and it is followed by two hidden layers,
which are comprised of 100 and 50 neurons, 
respectively. 
These dimensions were chosen after an optimisation process targeted at maximising the model accuracy.
Finally, the output layer is composed of two neurons, 
which capture the mean 
and the standard deviation 
of the power consumption,
as explained earlier.

\subsection{Training of the model}
The model is trained with the objective of reducing both the prediction error and its uncertainty. 
In particular, the training is considered successful, 
if the distribution outputted by the model for a given input 
matches the distribution of the power measurements in the data.

In terms of data management,
we split the available dataset related to 7760 \acp{AAU} into two parts:
a training set and a testing set. 
The training set contains data collected for 10 days, 
whereas the testing set contains the data collected for the 2 remaining days. 
The model training was performed by adopting the Adam version of the gradient descent algorithm~\cite{Goodfellow-et-al-2016}, 
and required 75 minutes to perform 1086 iterations.

\subsection{Model performance evaluation}
To assess the performance of the proposed \ac{ANN} model, 
we compared the estimated power consumption during the testing phase with the real measurements available in the data. 
Overall,
the model achieved a \ac{RMSE} of 25.02\,W, a \ac{MAE} of 12.21\,W and a remarkably low \ac{MAPE} of 6.55\,\% when estimating the power consumed by each \ac{AAU} in each hour of the testing period.

\begin{figure}[t!]
    \centering
    \includegraphics[clip,trim=0 0 0 1cm,width=8.5cm]{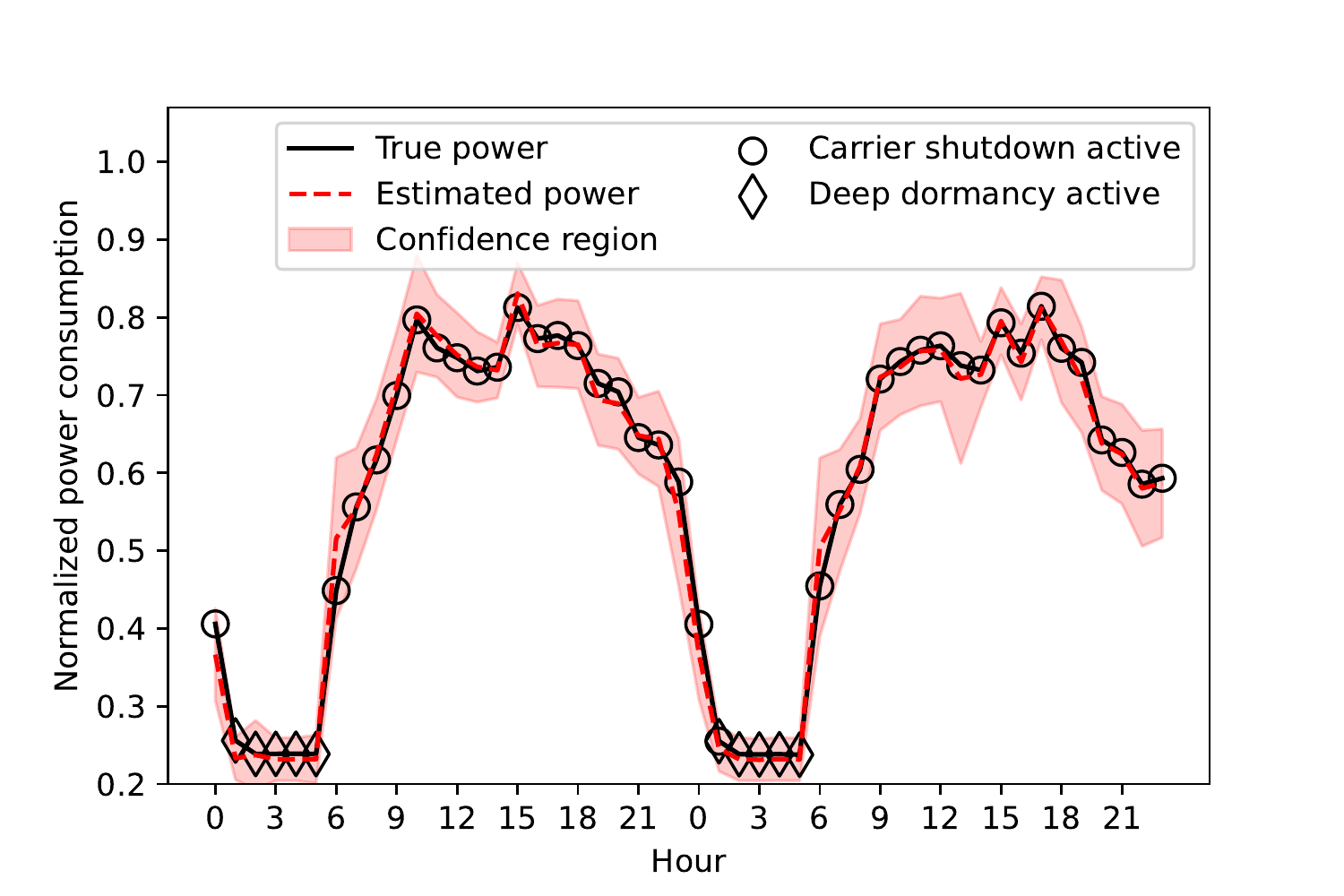}
    \caption{Hourly real and estimated normalised power consumption for an AAU doing carrier shutdown and deep dormancy.}
    \label{fig:MLdaily}
\end{figure}

To highlight the ability of the model to accurately estimate the power consumption when dynamic energy saving algorithms are activated,
Fig.~\ref{fig:MLdaily} shows an example of the real and estimated power consumption of a particular \ac{AAU},
which supported up to 6 carriers, 
and intensively used energy saving features,
during the 2 testing days. 
The confidence region is also reported, 
representing the interval in which the true power consumption is expected to fall with a 0.95 probability.
Note that we normalised the power consumption for privacy reasons.
From this figure,
it can be observed that the deep dormancy feature is activated during night hours 
(i.e., from 1am to 6am),
while the carrier shutdown algorithm is activated and intensively used during the rest of the day.
Note that an \ac{AAU} can only shutdown a carrier when its shutdown entry conditions are met,
and that such conditions mostly depend on traffic load, 
and are independently checked per carrier on a less than a minute basis.
Overall,
even in this highly dynamic activation/deactivation conditions,
the proposed model is able to estimate the power consumption of this \ac{AAU}
with high accuracy 
(i.e., \ac{RMSE} 14.43\,W, \ac{MAE} 9.5\,W, \ac{MAPE} 2.5\,\%). 

\begin{figure}[t!]
    \centering
    \includegraphics[clip,trim=0 0 0 1cm,width=8.5cm]{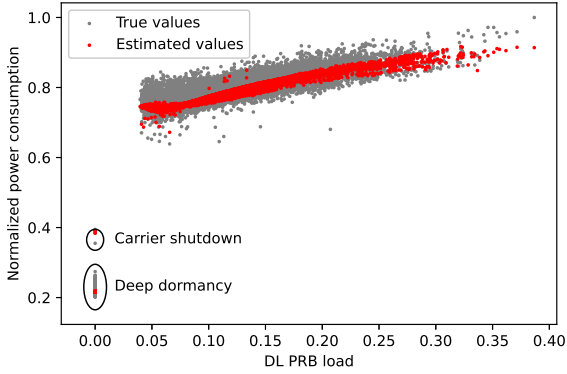}
    \caption{Normalised power consumption, estimated by the ANN model, versus \ac{AAU} \ac{DL} \ac{PRB} load for the selected \ac{AAU}.}
    \label{fig:ML_fit}
\end{figure}

To highlight the capability of the model to perform in a variety of deployment environments, 
Fig.~\ref{fig:ML_fit} also shows the real and estimated power consumption, 
not of a single \ac{AAU} as in Fig.~\ref{fig:MLdaily},
but for a popular \ac{AAU} type also supporting up to 6 carriers, 
which appears often in our dataset in different scenarios and city areas, 
with respect to the \ac{DL} \ac{PRB} load. 
For the sake of clarity, 
we would like to highlight here that the spread of the real and estimated values observed over the y-axis in this figure is motivated,
in addition to the noises introduced in Section~\ref{sec:output}, 
by the presence of multiple carriers deployed within this \ac{AAU} type, 
which are generally configured with different maximum transmit powers. 
As a result, there is not a biunivocal relation between the \ac{DL} \ac{PRB} load and the total transmit power (and thus neither with the power consumption).
From this figure,
it can be seen that this \ac{AAU} type achieves a 47\,\% and 70\,\% reduction in power consumption when doing carrier shutdown and deep dormancy, respectively. 
Importantly, 
even if this \ac{AAU} type was deployed in a heterogeneous set of scenarios,
the proposed model is able to accurately estimate the power consumption 
(i.e., \ac{RMSE} 18.25\,W, \ac{MAE} 14.48\,W, \ac{MAPE} 2.63\,\%).

\section{Analytical model}
\label{sec:math}
Although accurate and general,
the presented \ac{ML} model lacks tractability to drive  energy efficiency feature standardisation, development and/or optimisation.
To facilitate these tasks,
in this section,
based on the knowledge gathered from the previous \ac{ML} model,
we propose an analytically tractable \ac{5G} \ac{AAU} power consumption model,
which is easily interpretable and amicable to optimisation.

\subsection{Model description}
Our proposed \ac{5G} \ac{AAU} power consumption model,
which characterises the relationships between the key characteristics that play a major role on \ac{5G} \ac{AAU} power consumption, is mathematically formulated as  
\begin{equation}
    \begin{split}
    P_{\mathrm{AAU}} &= P_0 + P_{\rm BB} + \\ &+
    \underbrace{\sum_{t=1}^{T}  M_{\rm{av},t} D_{\rm{Tran},t}}_{P_{\mathrm{Tran}}} + \underbrace{M_{\rm{ac}} D_{\rm{PA}}}_{P_{\mathrm{PA}}}  +  \underbrace{\frac{1}{ \eta}  \sum_{c=1}^{C}  P_{\mathrm{TX},c}}_{P_{\mathrm{out}}}. 
    \end{split}
\end{equation}

In more details, 
the power, $P_{\mathrm{Tran}}$, consumed by the $t$-th transceiver in the \ac{AAU} is the product of the number of available \ac{RF} chains, $M_{\rm{av}}$, and the power consumed by each \ac{RF} chain, $D_{\rm{Tran},t}$.
The static power, $P_{\mathrm{PA}}$, consumed by the \acp{MCPA} is the product of the number of active \ac{RF} chains,  $M_{\rm{ac}}$, and the static power consumed by each \ac{MCPA}, $D_{\rm{PA}}$.
Recall that there is an \ac{MCPA} for each \ac{RF} chain spanning over the managed carriers.
Finally,
the power, $P_{\mathrm{out}}$, consumed to generate the transmit power required to transmit the data over the $c$-th CC is equal to the ratio of the transmit power in use at such CC, $P_{\mathrm{TX},c}$, to the efficiency of the \acp{MCPA} and antennas, $\eta$,
where the transmit power in use usually linearly increases with the number of \acp{PRB} utilised.

When symbol shutdown is activated, 
the \ac{AAU} switches off the \acp{MCPA}, 
and its power consumption is reduced to the sum of the baseline power consumption, $P_0$, the baseband processing power consumption, $P_{\rm{BB}}$, and the power consumed by the transceivers, $P_{\mathrm{Tran}}$ 
as they are not deactivated.

When channel shutdown is active, 
the \ac{AAU} reduces power consumption by limiting the multiplexing and beamforming capabilities of the cell, 
i.e., by limiting the number of active \acp{MCPA}.
This is realised in our model by decreasing the value of the variable, $M_{\rm{ac}}$, 
e.g., from 64 to 32 or 16.

When carrier shutdown is activated, 
the \acp{MCPA} and the transceivers are switched off.
Therefore, the power consumption is further reduced to the sum of the baseline power consumption, $P_0$, and the baseband processing power consumption, $P_{\rm{BB}}$. 
Finally, when deep dormancy is activated the circuitry for baseband processing is switched off, 
and  the \ac{AAU} power consumption is further reduced to the baseline power consumption, $P_0$.

The proposed model has a number of benefits,
not captured by other models in the literature,
which makes it a cornerstone for accurate 5G network energy efficiency standardisation, development and optimisation:
\begin{enumerate}
\item 
It allows to capture different multi-carrier architectures, 
i.e., intra-band contiguous, intra-band non-contiguous, and inter-band,  
where distinct carriers may or many not share the same transceiver;
\item 
It characterises realistic multi-carrier \ac{AAU} products with \acp{MCPA} and their intricate shutdown functioning,
where deactivating only a subset of the carriers in the \ac{AAU} does not lead to large energy savings, 
since the \acp{MCPA} must continue operating to support the active carriers;
\item 
It accounts for each of the state-of-the-art energy saving techniques, 
i.e., carrier shutdown, channel shutdown, symbol shutdown, and deep dormancy,
and can be easily extended to more.
\end{enumerate}

\subsection{Model fitting}

\begin{figure}[t!]
\vspace{+0.1cm}
    \centering
    \includegraphics[clip,trim=0 0 0 1.2cm,width=8.5cm]{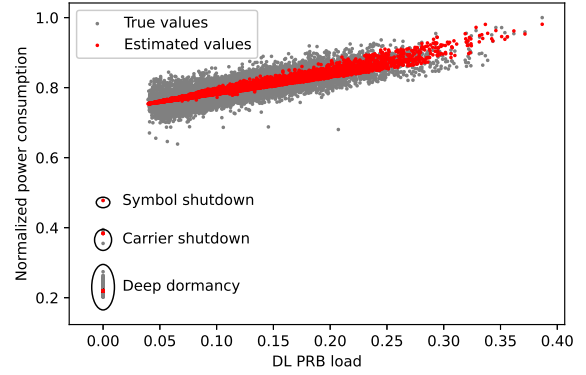}
    \caption{Normalised power consumption, estimated by the analytical model, versus AAU DL PRB load for a popular AAU.}
    \label{fig:math_fit}
\end{figure}

To analyse the proposed analytical model performance,
we have fitted its parameters,
for the popular \ac{AAU} type introduced in Sec.~\ref{sec:ml},
by using power consumption estimations performed through our \ac{ANN} model. 
Note that this approach to fitting,
not based on the data, 
but on \ac{ANN} model created trough the data,
allows to exploit the generalisation capabilities of our proposed \ac{ANN} model,
which can learn from other \ac{AAU} types,
and perform accurate power consumption estimations for traffic conditions not observed in the data of this \ac{AAU} type. 
In more detail, the analytical model parameters have been fitted on the generated data by iteratively solving a nonlinear least-squares regression problem.

The normalised values for the fitted parameters are 
$P_0=0.22$, $P_{\rm{BB}}=0.16$, $D_{\rm{Tran},1}=1.47\cdot 10^{-3}$, $D_{\rm{PA}}=3.81\cdot 10^{-3}$, $\eta=0.4$.
For completeness, 
let us note that the \ac{AAU} under study has $C=2$ \acp{CC} with $M_{\rm{av},1}=64$ RF chains. 

\subsection{Model performance evaluation}

Fig.~\ref{fig:math_fit} shows the normalised power consumed by the selected \ac{AAU} type for different values of the \ac{DL} \ac{PRB} load, 
observed in the dataset, 
and the values estimated by the fitted analytical model. 
The analytical model achieves a remarkable performance with RMSE 19.96\,W, MAE 16.50\,W and MAPE 2.67\,\%.
This estimation accuracy is close to that achieved by the \ac{ANN} model,
highlighting that the most relevant inputs to power consumption have been captured,
and the capability of the proposed analytical model to accurately model realistic \acp{AAU},
while considering the complex \ac{MCPA} structure and the existence of different energy saving modes.
A comparison of the accuracy performance reached by the \ac{ANN} and analytical models is reported in Table~\ref{tab:MLmath_comparison}.

From Fig.~\ref{fig:math_fit},
it can also be observed that, 
for this \ac{AAU} type, 
the activation of symbol shutdown provides a 34\,\%  power consumption saving w.r.t to the power consumption at zero load, 
while that of carrier shutdown results in larger savings,~47\,\%. 

It should be noted, however, that the lower power consumption achieved by carrier shutdown comes at the expense of an increased complexity in the network management.
Symbol shutdown operates locally ---and usually opportunistically--- in every cell at the time scale of hundreds of microseconds when no data needs to be transmitted, 
and thus it does not generally affect the user performance.
On the contrary, 
carrier shutdown strategies are adopted for longer time periods (from few minutes to few hours),
and coordinated across the network, 
as their activation requires/implies the redefinition of the network coverage and redistribution of its traffic,
i.e., user association.
Due to its complexity, 
if the carrier shutdown feature is not appropriately optimised,
energy savings may come at the expense of user experience.
Even worst, 
if the optimisation is performed with an inaccurate AAU power consumption model,
the energy saving gains may not even be there.

To illustrate this point, 
we have estimated the power consumption of the selected AAU under the same conditions over the 24 hours of a day with a state-of-the-art power consumption model~\cite{bjornson2015optimal}. 
Such model provided a 2.5$\times$ overestimation of the power consumption over the ground truth,
as it is not able to capture the multi-carrier architecture and the accurate impact of energy saving methods.
The error of our analytical model was less than 1\,\%.
This significant overestimation would lead to a suboptimal carrier shutdown configuration,
hindering energy savings,  
and shows how state-of-the-art models may fail to drive network energy efficiency optimisation.
Instead, 
the better accuracy of our proposed model indicates that it may be a more viable tool to drive the optimisation of greener 5G (and beyond) networks.

\begin{table}[]
\centering
\begin{tabular}{@{}llll@{}}
\toprule
Metric  & {Analytical} model & \ac{ML} model  & \ac{ML} model gain  \\ \midrule
RMSE    & 19.96\,W &   18.25\,W  & 8.6\,\%  \\
MAE     & 15.36\,W   &  14.48\,W & 5.7\,\%\\
MAPE    & 2.80\,\%    & 2.63\,\% & 6.1\,\% \\
\bottomrule
\end{tabular}
\caption{Comparison of accuracy performance achieved by the ANN model and the analytical model for a popular AAU type.}
\label{tab:MLmath_comparison}
\end{table}

\section{Conclusions}
\label{sec:conclusions}
In this paper, 
we presented a novel power consumption model for realistic \ac{5G} \acp{AAU}, 
which builds on a large data collection campaign. 
At first, we proposed an \ac{ANN} architecture,
which allows modelling multiple types of \ac{AAU} and different configurations. 
The discussed results highlighted that the designed \ac{ANN} architecture is able to provide high accuracy. 
In a second stage, 
we exploited the knowledge gathered by the \ac{ANN} method to derive a novel and realistic but analytically tractable \ac{5G} \ac{AAU} power consumption model. 
We demonstrated that such analytical model reaches accuracy close to the one of the \ac{ML} model for a widely used type of \ac{AAU}. 
Notably, 
when compared to a state-of-the-art model under the same conditions, 
the proposed one was shown to be around 150\,\% more accurate,
as it is able of precisely capturing the \ac{MCPA} architecture and the benefits of shutdown approaches.
Importantly, due to its fundamental nature, the proposed methodology can be adopted to model other types of \ac{AAU} deployed in different multi-vendor networks.
We thus believe that this model 
is a valuable contribution to both industry and the research community working on wireless network energy efficiency,
and its optimisation, and can be of use in the current \ac{3GPP} \ac{NR} Release 18 work on network energy efficiency.


\bibliographystyle{IEEEtran}
\bibliography{reference.bib}
\vspace{-1cm}
\begin{IEEEbiographynophoto}
    {Nicola Piovesan} (Member, IEEE)
is a Senior Researcher with Huawei Technologies, France. His research interests include energy sustainability, energy efficiency optimization, and \ac{ML} in wireless communication systems.
\end{IEEEbiographynophoto}
\vspace{-1.1cm}
\begin{IEEEbiographynophoto}
    {David Lopez-Perez}  (Senior Member, IEEE)
is an Expert and a Technical Lead with Huawei Technologies, France. 
His interests are on both cellular and Wi-Fi networks, network performance analysis, network planning and optimization, as well as technology and feature development.
\end{IEEEbiographynophoto}
\vspace{-1.1cm}
\begin{IEEEbiographynophoto}
    {Antonio De Domenico} (Member, IEEE)
is a Senior Researcher with Huawei Technologies, France. His research interests include heterogeneous wireless networks, \ac{ML}, and green communications. 
\end{IEEEbiographynophoto}
\vspace{-1.1cm}
\begin{IEEEbiographynophoto}
    {Harvey Bao} is a Senior Researcher with Huawei Technologies, France.  His research interests include new technologies for wireless networks and AI-driven network modelling and optimisation.
\end{IEEEbiographynophoto}
\vspace{-1.1cm}
\begin{IEEEbiographynophoto}
    {Xinli Geng}
    is a Principal Engineer 
    with Huawei Technologies, China. His current research interests include wireless networks optimisation, green communications, data-driven network modelling and AI-related technologies.
\end{IEEEbiographynophoto}
\vspace{-1.1cm}
\begin{IEEEbiographynophoto}
    {M\'erouane Debbah}
    (Fellow, IEEE) is Chief Research Officer at the Technology Innovation Institute in Abu Dhabi. From 2014 to 2021, he was Vice-President of the Huawei France Research Center, where he was jointly the director of the Mathematical and Algorithmic Sciences Lab as well as the director of the Lagrange Mathematical and Computing Research Center. 
\end{IEEEbiographynophoto}
\begin{acronym}[AAAAAAAAA]

 \acro{2D}{two-dimensional}
 \acro{3D}{three-dimensional}
 \acro{3G}{third generation}
 \acro{3GPP}{third generation partnership project}
 \acro{4G}{fourth generation}
 \acro{5G}{fifth generation}
 \acro{5GC}{5G Core Network}
 \acro{6G}{sixth generation}
 \acro{AAA}{authentication, authorisation and accounting}
 \acro{AAU}{active antenna unit}
 \acro{ABS}{almost blank subframe}
 \acro{AC}{alternating current}
 \acro{ACIR}{adjacent channel interference rejection ratio}
 \acro{ACK}{acknowledgment}
 \acro{ACL}{allowed CSG list}
 \acro{ACLR}{adjacent channel leakage ratio}
 \acro{ACPR}{adjacent channel power ratio}
 \acro{ACS}{adjacent channel selectivity}
 \acro{ADC}{analog-to-digital converter}
 \acro{ADSL}{asymmetric digital subscriber line}
 \acro{AEE}{area energy efficiency}
 \acro{AF}{amplify-and-forward}
 \acro{AGCH}{access grant channel}
 \acro{AGG}{aggressor cell}
 \acro{AH}{authentication header}
 \acro{AI}{artificial intelligence}
 \acro{AKA}{authentication and key agreement}
 \acro{AMC}{adaptive modulation and coding}
 \acro{ANN}{artificial neural network}
 \acro{ANR}{automatic neighbour relation}
 \acro{AoA}{angle of arrival}
 \acro{AoD}{angle of departure}
 \acro{APC}{area power consumption}
 \acro{API}{application programming interface}
 \acro{APP}{a posteriori probability}
 \acro{AR}{augmented reality}
 \acro{ARIMA}{autoregressive integrated moving average}
 \acro{ARQ}{automatic repeat request}
 \acro{AS}{access stratum}
 \acro{ASE}{area spectral efficiency}
 \acro{ASM}{advanced sleep mode}
 \acro{ASN}{access service network}
 \acro{ATM}{asynchronous transfer mode}
 \acro{ATSC}{Advanced Television Systems Committee}
 \acro{AUC}{authentication centre}
 \acro{AWGN}{additive white gaussian noise}
 \acro{BB}{baseband}
 \acro{BBU}{baseband unit}
 \acro{BCCH}{broadcast control channel}
 \acro{BCH}{broadcast channel}
 \acro{BCJR}{Bahl-Cocke-Jelinek-Raviv} 
 \acro{BE}{best effort}
 \acro{BER}{bit error rate}
 \acro{BLER}{block error rate}
 \acro{BPSK}{binary phase-shift keying}
 \acro{BR}{bit rate}
 \acro{BS}{base station}
 \acro{BSC}{base station controller}
 \acro{BSIC}{base station identity code}
 \acro{BSP}{binary space partitioning}
 \acro{BSS}{blind source separation}
 \acro{BTS}{base transceiver station}
 \acro{BWP}{Bandwidth Part}
 \acro{CA}{carrier aggregation}
 \acro{CAC}{call admission control}
 \acro{CaCo}{carrier component}
 \acro{CAPEX}{capital expenditure}
 \acro{capex}{capital expenses}
 \acro{CAS}{cluster angular spread}
 \acro{CATV}{community antenna television}
 \acro{CAZAC}{constant amplitude zero auto-correlation}
 \acro{CC}{component carrier}
 \acro{CCCH}{common control channel}
 \acro{CCDF}{complementary cumulative distribution function}
 \acro{CCE}{control channel element}
 \acro{CCO}{coverage and capacity optimisation}
 \acro{CCPCH}{common control physical channel}
 \acro{CCRS}{coordinated and cooperative relay system}
 \acro{CCTrCH}{coded composite transport channel}
 \acro{CDF}{cumulative distribution function}
 \acro{CDMA}{code division multiple access}
 \acro{CDS}{cluster delay spread}
 \acro{CESM}{capacity effective SINR mapping}
 \acro{CO$_{2e}$}{carbon dioxide equivalent}
 \acro{CFI}{control format indicator}
 \acro{CFL}{Courant-Friedrichs-Lewy}
 \acro{CGI}{cell global identity}
 \acro{CID}{connection identifier}
 \acro{CIF}{carrier indicator field}
 \acro{CIO}{cell individual offset}
 \acro{CIR}{channel impulse response}
 \acro{CNN}{Convolutional Neural Network}
 \acro{CMF}{cumulative mass function}
 \acro{C-MIMO}{cooperative MIMO}
 \acro{CN}{core network}
 \acro{COC}{cell outage compensation}
 \acro{COD}{cell outage detection}
 \acro{CoMP}{coordinated multi-point}
 \acro{ConvLSTM}{Convolutional LSTM}
 \acro{CP}{cycle prefix}
 \acro{CPC}{cognitive pilot channel}
 \acro{CPCH}{common packet channel}
 \acro{CPE}{customer premises equipment}
 \acro{CPICH}{common pilot channel}
 \acro{CPRI}{common public radio interface}
 \acro{CPU}{central processing unit}
 \acro{CQI}{channel quality indicator}
 \acro{CR}{cognitive radio}
 \acro{CRAN}{centralized radio access network} 
 \acro{C-RAN}{cloud radio access network} 
 \acro{CRC}{cyclic redundancy check}
 \acro{CRE}{cell range expansion}
 \acro{C-RNTI}{cell radio network temporary identifier}
 \acro{CRP}{cell re-selection parameter}
 \acro{CRS}{cell-specific reference symbol}
 \acro{CRT}{cell re-selection threshold}
 \acro{CSCC}{common spectrum coordination channel}
 \acro{CSG ID}{closed subscriber group ID}
 \acro{CSG}{closed subscriber group}
 \acro{CSI}{channel state information}
 \acro{CSIR}{receiver-side channel state information}
 \acro{CSI-RS}{channel state information-reference signals}
 \acro{CSO}{cell selection offset}
 \acro{CTCH}{common traffic channel}
 \acro{CTS}{clear-to-send} 
 \acro{CU}{central unit}
 \acro{CV}{cross-validation}
 \acro{CWiND}{Centre for Wireless Network Design}
 \acro{D2D}{device to device}
 \acro{DAB}{digital audio broadcasting}
 \acro{DAC}{digital-to-analog converter}
 \acro{DAS}{distributed antenna system}
 \acro{dB}{decibel}
 \acro{dBi}{isotropic-decibel}
 \acro{DC}{direct current}
 \acro{DCCH}{dedicated control channel}
 \acro{DCF}{decode-and-forward}
 \acro{DCH}{dedicated channel}
 \acro{DC-HSPA}{dual-carrier high speed packet access}
 \acro{DCI}{downlink control information}
 \acro{DCM}{directional channel model}
 \acro{DCP}{dirty-paper coding}
 \acro{DCS}{digital communication system}
 \acro{DECT}{digital enhanced cordless telecommunication}
 \acro{DeNB}{donor eNodeB}
 \acro{DFP}{dynamic frequency planning}
 \acro{DFS}{dynamic frequency selection}
 \acro{DFT}{discrete Fourier transform}
 \acro{DFTS}{discrete Fourier transform spread}
 \acro{DHCP}{dynamic host control protocol}
 \acro{DL}{downlink}
 \acro{DMC}{dense multi-path components}
 \acro{DMF}{demodulate-and-forward}
 \acro{DMT}{diversity and multiplexing tradeoff}
  \acro{DNN}{deep neural network} 
 \acro{DoA}{direction-of-arrival}
 \acro{DoD}{direction-of-departure}
 \acro{DoS}{denial of service}
 \acro{DPCCH}{dedicated physical control channel}
 \acro{DPDCH}{dedicated physical data channel}
 \acro{D-QDCR}{distributed QoS-based dynamic channel reservation}
 \acro{DQL}{deep Q-learning}
  \acro{DRAN}{distributed radio access network}
 \acro{DRS}{discovery reference signal}
 \acro{DRL}{deep reinforcement learning}
 \acro{DRX}{discontinuous reception}
 \acro{DS}{down stream}
 \acro{DSA}{dynamic spectrum access}
 \acro{DSCH}{downlink shared channel}
 \acro{DSL}{digital subscriber line}
 \acro{DSLAM}{digital subscriber line access multiplexer}
 \acro{DSP}{digital signal processor}
 \acro{DT}{decision tree}
 \acro{DTCH}{dedicated traffic channel}
 \acro{DTX}{discontinuous transmission}
   \acro{DU}{distributed unit}
 \acro{DVB}{digital video broadcasting}
 \acro{DXF}{drawing interchange format}
 \acro{E2E}{end-to-end}
 \acro{EAGCH}{enhanced uplink absolute grant channel}
 \acro{EA}{evolutionary algorithm}
 \acro{EAP}{extensible authentication protocol}
 \acro{EC}{evolutionary computing}
 \acro{ECGI}{evolved cell global identifier}
 \acro{ECR}{energy consumption ratio}
 \acro{ECRM}{effective code rate map}
 \acro{EDCH}{enhanced dedicated channel}
 \acro{EE}{energy efficiency}
 \acro{EESM}{exponential effective SINR mapping}
 \acro{EF}{estimate-and-forward}
 \acro{EGC}{equal gain combining}
 \acro{EHICH}{EDCH HARQ indicator channel}
 \acro{eICIC}{enhanced intercell interference coordination}
 \acro{EIR}{equipment identity register}
 \acro{EIRP}{effective isotropic radiated power}
 \acro{ELF}{evolutionary learning of fuzzy rules}
 \acro{eMBB}{enhanced mobile broadband}
  \acro{EMR}{Electromagnetic-Radiation}
 \acro{EMS}{enhanced messaging service}
 \acro{eNB}{evolved NodeB}
 \acro{eNodeB}{evolved NodeB}
 \acro{EoA}{elevation of arrival}
 \acro{EoD}{elevation of departure}
 \acro{EPB}{equal path-loss boundary}
 \acro{EPC}{evolved packet core}
 \acro{EPDCCH}{enhanced physical downlink control channel}
 \acro{EPLMN}{equivalent PLMN}
 \acro{EPS}{evolved packet system}
 \acro{ERAB}{eUTRAN radio access bearer}
 \acro{ERGC}{enhanced uplink relative grant channel}
 \acro{ERTPS}{extended real time polling service}
 \acro{ESB}{equal downlink receive signal strength boundary}
 \acro{ESF}{even subframe}
 \acro{ESP}{encapsulating security payload}
 \acro{ETSI}{European Telecommunications Standards Institute}
 \acro{E-UTRA}{evolved UTRA}
 \acro{EU}{European Union}
 \acro{EUTRAN}{evolved UTRAN}
 \acro{EVDO}{evolution-data optimised}
 \acro{FACCH}{fast associated control channel}
 \acro{FACH}{forward access channel}
 \acro{FAP}{femtocell access point}
 \acro{FARL}{fuzzy assisted reinforcement learning}
 \acro{FCC}{Federal Communications Commission}
 \acro{FCCH}{frequency-correlation channel}
 \acro{FCFS}{first-come first-served}
 \acro{FCH}{frame control header}
 \acro{FCI}{failure cell ID}
 \acro{FD}{frequency-domain}
 \acro{FDD}{frequency division duplexing}
 \acro{FDM}{frequency division multiplexing}
 \acro{FDMA}{frequency division multiple access}
 \acro{FDTD}{finite-difference time-domain}
 \acro{FE}{front-end}
 \acro{FeMBMS}{further evolved multimedia broadcast multicast service}
 \acro{FER}{frame error rate}
 \acro{FFR}{fractional frequency reuse}
 \acro{FFRS}{fractional frequency reuse scheme}
 \acro{FFT}{fast Fourier transform}
 \acro{FFU}{flexible frequency usage}
 \acro{FGW}{femtocell gateway}
 \acro{FIFO}{first-in first-out}
 \acro{FIS}{fuzzy inference system}
 \acro{FMC}{fixed mobile convergence}
 \acro{FPC}{fractional power control}
 \acro{FPGA}{field-programmable gate array}
 \acro{FRS}{frequency reuse scheme}
 \acro{FTP}{file transfer protocol}
 \acro{FTTx}{fiber to the x}
 \acro{FUSC}{full usage of subchannels}
 \acro{GA}{genetic algorithm}
 \acro{GAN} {generic access network}
 \acro{GANC}{generic access network controller}
 \acro{GBR}{guaranteed bitrate}
 \acro{GCI}{global cell identity}
 \acro{GCN}{Graph convolutional network}
 \acro{GERAN}{GSM edge radio access network}
 \acro{GGSN}{gateway GPRS support node}
 \acro{GHG}{greenhouse gas}
 \acro{GMSC}{gateway mobile switching centre}
 \acro{gNB}{next generation NodeB}
 \acro{GNN}{Graph Neural Network}
 \acro{GNSS}{global navigation satellite system}
 \acro{GP}{genetic programming}
 \acro{GPON}{Gigabit passive optical network}
 \acro{GPP}{general purpose processor}
 \acro{GPRS}{general packet radio service}
 \acro{GPS}{global positioning system}
 \acro{GPU}{graphics processing unit}
 \acro{GRU}{gated recurrent unit}
 \acro{GSCM}{geometry-based stochastic channel models}
 \acro{GSM}{global system for mobile communication}
 \acro{GTD}{geometry theory of diffraction}
 \acro{GTP}{GPRS tunnel protocol}
 \acro{GTP-U}{GPRS tunnel protocol - user plane}
 \acro{HA}{historical average}
 \acro{HARQ}{hybrid automatic repeat request}
 \acro{HBS}{home base station}
 \acro{HCN}{heterogeneous cellular network}
 \acro{HCS}{hierarchical cell structure}
  \acro{HD}{high definition}
 \acro{HDFP}{horizontal dynamic frequency planning}
 \acro{HeNB}{home eNodeB}
 \acro{HeNodeB}{home eNodeB}
 \acro{HetNet}{heterogeneous network}
 \acro{HiFi}{high fidelity}
 \acro{HII}{high interference indicator}
 \acro{HLR}{home location register}
 \acro{HNB}{home NodeB}
 \acro{HNBAP}{home NodeB application protocol}
 \acro{HNBGW}{home NodeB gateway}
 \acro{HNodeB}{home NodeB}
 \acro{HO}{handover}
 \acro{HOF}{handover failure}
 \acro{HOM}{handover hysteresis margin}
 \acro{HPBW}{half power beam width}
 \acro{HPLMN}{home PLMN}
 \acro{HPPP}{homogeneous Poison point process}
 \acro{HRD}{horizontal reflection diffraction}
 \acro{HSB}{hot spot boundary}
 \acro{HSDPA}{high speed downlink packet access}
 \acro{HSDSCH}{high-speed DSCH}
 \acro{HSPA}{high speed packet access}
 \acro{HSS}{home subscriber server}
 \acro{HSUPA}{high speed uplink packet access}
 \acro{HUA}{home user agent}
 \acro{HUE}{home user equipment}
 \acro{HVAC}{heating, ventilating, and air conditioning}
 \acro{HW}{Holt-Winters}
 \acro{IC}{interference cancellation}
 \acro{ICI}{inter-carrier interference}
 \acro{ICIC}{intercell interference coordination}
 \acro{ICNIRP}{International Commission on Non-Ionising Radiation Protection}
 \acro{ICS}{IMS centralised service}
 \acro{ICT}{information and communication technology}
 \acro{ID}{identifier}
 \acro{IDFT}{inverse discrete Fourier transform}
 \acro{IE}{information element}
 \acro{IEEE}{Institute of Electrical and Electronics Engineers}
 \acro{IETF}{Internet engineering task force}
 \acro{IFA}{Inverted-F-antennas}
 \acro{IFFT}{inverse fast Fourier transform}
 \acro{i.i.d.}{independent and identical distributed}
 \acro{IIR}{infinite impulse response}
 \acro{IKE}{Internet key exchange}
 \acro{IKEv2}{Internet key exchange version 2}
 \acro{ILP}{integer linear programming}
 \acro{IMEI}{international mobile equipment identity}
 \acro{IMS}{IP multimedia subsystem}
 \acro{IMSI}{international mobile subscriber identity}
 \acro{IMT}{international mobile telecommunications}
 \acro{INH}{indoor hotspot}
 \acro{IOI}{interference overload indicator}
 \acro{IoT}{Internet of things}
 \acro{IP}{Internet protocol}
 \acro{IPSEC}{Internet protocol security}
 \acro{IR}{incremental redundancy}
 \acro{IRC}{interference rejection combining}
 \acro{ISD}{inter site distance}
 \acro{ISI}{inter symbol interference}
 \acro{ITU}{International Telecommunication Union}
 \acro{Iub}{UMTS interface between RNC and NodeB}
 \acro{IWF}{IMS interworking function}
 \acro{JFI}{Jain's fairness index}
 \acro{KPI}{key performance indicator}
 \acro{KNN}{k-nearest neighbours}
 \acro{L1}{layer one}
 \acro{L2}{layer two}
 \acro{L3}{layer three}
 \acro{LA}{location area}
 \acro{LAA}{licensed Assisted Access}
 \acro{LAC}{location area code}
 \acro{LAI}{location area identity}
 \acro{LAU}{location area update}
 \acro{LDA}{linear discriminant analysis} 
 \acro{LIDAR}{laser imaging detection and ranging}
 \acro{LLR}{log-likelihood ratio}
 \acro{LLS}{link-level simulation}
 \acro{LMDS}{local multipoint distribution service}
 \acro{LMMSE}{linear minimum mean-square-error}
 \acro{LoS}{line-of-sight}
 \acro{LPC}{logical PDCCH candidate}
 \acro{LPN}{low power node}
 \acro{LR}{likelihood ratio}
 \acro{LSAS}{large-scale antenna system}
 \acro{LSP}{large-scale parameter}
 \acro{LSTM}{long short term memory cell}
 \acro{LTE/SAE}{long term evolution/system architecture evolution}
 \acro{LTE}{long term evolution}
 \acro{LTE-A}{long term evolution advanced}
 \acro{LUT}{look up table}
 \acro{MAC}{medium access control}
 \acro{MaCe}{macro cell}
  \acro{MAE}{mean absolute error}
 \acro{MAP}{media access protocol}
 \acro{MAPE}{mean absolute percentage error}
 \acro{MAXI}{maximum insertion}
 \acro{MAXR}{maximum removal}
 \acro{MBMS}{multicast broadcast multimedia service} 
 \acro{MBS}{macrocell base station}
 \acro{MBSFN}{multicast-broadcast single-frequency network}
 \acro{MC}{modulation and coding}
 \acro{MCB}{main circuit board}
 \acro{MCM}{multi-carrier modulation}
 \acro{MCP}{multi-cell processing}
 \acro{MCPA}{multi-carrier power amplifier}
 \acro{MCS}{modulation and coding scheme}
 \acro{MCSR}{multi-carrier soft reuse}
 \acro{MDAF}{management data analytics function}
 \acro{MDP}{markov decision process }
 \acro{MDT}{minimisation of drive tests}
 \acro{MEA}{multi-element antenna}
 \acro{MeNodeB}{Master eNodeB}
 \acro{MGW}{media gateway}
 \acro{MIB}{master information block}
 \acro{MIC}{mean instantaneous capacity}
 \acro{MIESM}{mutual information effective SINR mapping}
 \acro{MIMO}{multiple-input multiple-output}
 \acro{MINI}{minimum insertion}
 \acro{MINR}{minimum removal}
 \acro{MIP}{mixed integer program}
 \acro{MISO}{multiple-input single-output}
 \acro{ML}{machine learning}
 \acro{MLB}{mobility load balancing}
 \acro{MLB}{mobility load balancing}
 \acro{MM}{mobility management}
 \acro{MME}{mobility management entity}
 \acro{mMIMO}{massive multiple-input multiple-output}
 \acro{MMSE}{minimum mean square error}
 \acro{mMTC}{massive machine type communication}
 \acro{MNC}{mobile network code}
 \acro{MNO}{mobile network operator}
 \acro{MOS}{mean opinion score}
 \acro{MPC}{multi-path component}
 \acro{MR}{measurement report}
 \acro{MRC}{maximal ratio combining}
 \acro{MR-FDPF}{multi-resolution frequency-domain parflow}
 \acro{MRO}{mobility robustness optimisation}
 \acro{MRT}{Maximum Ratio Transmission}
 \acro{MS}{mobile station}
 \acro{MSC}{mobile switching centre}
 \acro{MSE}{mean square error}
 \acro{MSISDN}{mobile subscriber integrated services digital network number}
 \acro{MUE}{macrocell user equipment}
 \acro{MU-MIMO}{multi-user MIMO}
 \acro{MVNO}{mobile virtual network operators}
 \acro{NACK}{negative acknowledgment}
 \acro{NAS}{non access stratum}
 \acro{NAV}{network allocation vector}
 \acro{NB}{Naive Bayes}   
 \acro{NCL}{neighbour cell list}
 \acro{NEE}{network energy efficiency}
  \acro{NF}{Network Function}
 \acro{NFV}{Network Functions Virtualization}
 \acro{NG}{next generation}
 \acro{NGMN}{next generation mobile networks}
 \acro{NG-RAN}{next generation radio access network} 
 \acro{NIR}{non ionisation radiation}
 \acro{NLoS}{non-line-of-sight}
 \acro{NN}{nearest neighbour} 
 \acro{NR}{new radio}
 \acro{NRTPS}{non-real-time polling service}
 \acro{NSS}{network switching subsystem}
 \acro{NTP}{network time protocol}
 \acro{NWG}{network working group}
 \acro{NWDAF}{network data analytics function} 
 \acro{OA}{open access}
 \acro{OAM}{operation, administration and maintenance}
 \acro{OC}{optimum combining}
 \acro{OCXO}{oven controlled oscillator}
 \acro{ODA}{omdi-directional antenna} 
 \acro{ODU}{optical distribution unit}
 \acro{OFDM}{orthogonal frequency division multiplexing}
 \acro{OFDMA}{orthogonal frequency division multiple access}
 \acro{OFS}{orthogonally-filled subframe}
 \acro{OLT}{optical line termination}
 \acro{ONT}{optical network terminal}
 \acro{OPEX}{operational expenditure}
 \acro{OSF}{odd subframe}
 \acro{OSI}{open systems interconnection}
 \acro{OSS}{operation support subsystem}
 \acro{OTT}{over the top}
 \acro{P2MP}{point to multi-point}
 \acro{P2P}{point to point}
 \acro{PAPR}{peak-to-average power ratio}
 \acro{PA}{power amplifier}
 \acro{PBCH}{physical broadcast channel}
 \acro{PC}{power control}
 \acro{PCB}{printed circuit board}
 \acro{PCC}{primary carrier component}
 \acro{PCCH}{paging control channel}
 \acro{PCCPCH}{primary common control physical channel}
 \acro{PCell}{primary cell}
 \acro{PCFICH}{physical control format indicator channel}
 \acro{PCH}{paging channel}
 \acro{PCI}{physical layer cell identity}
 \acro{PCPICH}{primary common pilot channel}
 \acro{PCPPH}{physical common packet channel}
 \acro{PDCCH}{physical downlink control channel}
 \acro{PDCP}{packet data convergence protocol}
 \acro{PDF}{probability density function}
 \acro{PDSCH}{physical downlink shared channel}
 \acro{PDU}{packet data unit}
 \acro{PeNB}{pico eNodeB}
 \acro{PeNodeB}{pico eNodeB}
 \acro{PF}{proportional fair}
 \acro{PGW}{packet data network gateway}
 \acro{PGFL}{probability generating functional}
 \acro{PhD}{doctor of philosophy}
 \acro{PHICH}{physical HARQ indicator channel}
 \acro{PHY}{physical}
 \acro{PIC}{parallel interference cancellation}
 \acro{PKI}{public key infrastructure}
 \acro{PL}{path loss}
 \acro{PMI}{precoding-matrix indicator}
 \acro{PLMN ID}{public land mobile network identity}
 \acro{PLMN}{public land mobile network}
 \acro{PML}{perfectly matched layer}
 \acro{PMF}{probability mass function}
 \acro{PMP}{point to multi-point}
 \acro{PN}{pseudorandom noise}
 \acro{POI}{point of interest}
 \acro{PON}{passive optical network}
 \acro{POP}{point of presence}
 \acro{PP}{point process}
 \acro{PPP}{Poisson point process}
 \acro{PPT}{PCI planning tools}
 \acro{PRACH}{physical random access channel}
 \acro{PRB}{physical resource block}
 \acro{PSC}{primary scrambling code}
 \acro{PSD}{power spectral density}
 \acro{PSS}{primary synchronisation channel}
 \acro{PSTN}{public switched telephone network}
 \acro{PTP}{point to point}
 \acro{PUCCH}{Physical Uplink Control Channel}
 \acro{PUE}{picocell user equipment}
 \acro{PUSC}{partial usage of subchannels}
 \acro{PUSCH}{physical uplink shared channel}
 \acro{QAM}{quadrature amplitude modulation}
 \acro{QCI}{QoS class identifier}
 \acro{QoE}{quality of experience}
 \acro{QoS}{quality of service}
 \acro{QPSK}{quadrature phase-shift keying}
 \acro{RAB}{radio access bearer}
 \acro{RACH}{random access channel}
 \acro{RADIUS}{remote authentication dial-in user services}
 \acro{RAN}{radio access network}
 \acro{RANAP}{radio access network application part}
 \acro{RAT}{radio access technology}
 \acro{RAU}{remote antenna unit}
 \acro{RAXN}{relay-aided x network}
 \acro{RB}{resource block}
 \acro{RCI}{re-establish cell id}
 \acro{RE}{resource efficiency}
 \acro{REB}{range expansion bias}
 \acro{REG}{resource element group}
 \acro{RF}{radio frequency}
  \acro{RFID}{radio frequency identification}
 \acro{RFP}{radio frequency planning}
 \acro{RI}{rank indicator}
 \acro{RL}{reinforcement learning}
 \acro{RLC}{radio link control}
 \acro{RLF}{radio link failure}
 \acro{RLM}{radio link monitoring}
 \acro{RMA}{rural macrocell}
 \acro{RMS}{root mean square}
 \acro{RMSE}{root mean square error}
 \acro{RN}{relay node}
 \acro{RNC}{radio network controller}
 \acro{RNL}{radio network layer}
 \acro{RNN}{recurrent neural network}
 \acro{RNP}{radio network planning}
 \acro{RNS}{radio network subsystem}
 \acro{RNTI}{radio network temporary identifier}
 \acro{RNTP}{relative narrowband transmit power}
 \acro{RPLMN}{registered PLMN}
 \acro{RPSF}{reduced-power subframes}
 \acro{RR}{round robin}
 \acro{RRC}{radio resource control}
 \acro{RRH}{remote radio head}
 \acro{RRM}{radio resource management}
 \acro{RS}{reference signal}
 \acro{RSC}{recursive systematic convolutional}
 \acro{RS-CS}{resource-specific cell-selection}
 \acro{RSQ}{reference signal quality}
 \acro{RSRP}{reference signal received power}
 \acro{RSRQ}{reference signal received quality}
 \acro{RSS}{reference signal strength}
 \acro{RSSI}{receive signal strength indicator}
 \acro{RTP}{real time transport}
 \acro{RTPS}{real-time polling service}
 \acro{RTS}{request-to-send}
 \acro{RTT}{round trip time}
  \acro{RU}{remote unit}
  \acro{RV}{random variable}
 \acro{RX}{receive}
 \acro{S1-AP}{s1 application protocol}
 \acro{S1-MME}{s1 for the control plane}
 \acro{S1-U}{s1 for the user plane}
 \acro{SA}{simulated annealing}
 \acro{SACCH}{slow associated control channel}
 \acro{SAE}{system architecture evolution}
 \acro{SAEGW}{system architecture evolution gateway}
 \acro{SAIC}{single antenna interference cancellation}
 \acro{SAP}{service access point}
 \acro{SAR}{specific absorption rate}
 \acro{SARIMA}{seasonal autoregressive integrated moving average}
 \acro{SAS}{spectrum allocation server}
 \acro{SBS}{super base station}
 \acro{SCC}{standards coordinating committee}
 \acro{SCCPCH}{secondary common control physical channel}
 \acro{SCell}{secondary cell}
 \acro{SCFDMA}{single carrier FDMA}
 \acro{SCH}{synchronisation channel}
 \acro{SCM}{spatial channel model}
 \acro{SCN}{small cell network}
 \acro{SCOFDM}{single carrier orthogonal frequency division multiplexing}
 \acro{SCP}{single cell processing}
 \acro{SCTP}{stream control transmission protocol}
 \acro{SDCCH}{standalone dedicated control channel}
 \acro{SDMA}{space-division multiple-access}
  \acro{SDO}{standard development organization}
 \acro{SDR}{software defined radio}
 \acro{SDU}{service data unit}
 \acro{SE}{spectral efficiency}
 \acro{SeNodeB}{secondary eNodeB}
 \acro{Seq2Seq}{Sequence-to-sequence}
 \acro{SFBC}{space frequency block coding}
 \acro{SFID}{service flow ID}
 \acro{SG}{signalling gateway}
 \acro{SGSN}{serving GPRS support node}
 \acro{SGW}{serving gateway}
 \acro{SI}{system information}
 \acro{SIB}{system information block}
 \acro{SIB1}{systeminformationblocktype1}
 \acro{SIB4}{systeminformationblocktype4}
 \acro{SIC}{successive interference cancellation}
 \acro{SIGTRAN}{signalling transport}
 \acro{SIM}{subscriber identity module}
 \acro{SIMO}{single input multiple output}
 \acro{SINR}{signal to interference plus noise ratio}
 \acro{SIP}{session initiated protocol}
 \acro{SIR}{signal to interference ratio}
 \acro{SISO}{single input single output}
 \acro{SLAC}{stochastic local area channel}
 \acro{SLL}{secondary lobe level}
 \acro{SLNR}{signal to leakage interference and noise ratio}
 \acro{SLS}{system-level simulation}
 \acro{SMAPE}{symmetric mean absolute percentage error}
 \acro{SMB}{small and medium-sized businesses}
 \acro{SmCe}{small cell}
 \acro{SMS}{short message service}
 \acro{SN}{serial number}
 \acro{SNMP}{simple network management protocol}
 \acro{SNR}{signal to noise ratio}
 \acro{SOCP}{second-order cone programming}
 \acro{SOHO}{small office/home office}
 \acro{SON}{self-organising network}
 \acro{son}{self-organising networks}
 \acro{SOT}{saving of transmissions}
 \acro{SPS}{spectrum policy server}
 \acro{SRS}{sounding reference signals}
 \acro{SS}{synchronization signal}
 \acro{SSB}{synchronisation signal block}
 \acro{SSL}{secure socket layer}
 \acro{SSMA}{spread spectrum multiple access}
 \acro{SSS}{secondary synchronisation channel}
 \acro{ST}{spatio temporal}
 \acro{STA}{steepest ascent}
 \acro{STBC}{space-time block coding}
 \acro{SUI}{stanford university interim}
 \acro{SVR}{support vector regression}
 \acro{TA}{timing advance}
 \acro{TAC}{tracking area code}
 \acro{TAI}{tracking area identity}
 \acro{TAS}{transmit antenna selection}
 \acro{TAU}{tracking area update}
 \acro{TCH}{traffic channel}
 \acro{TCO}{total cost of ownership}
 \acro{TCP}{transmission control protocol}
 \acro{TCXO}{temperature controlled oscillator}
 \acro{TD}{temporal difference}
 \acro{TDD}{time division duplexing}
 \acro{TDM}{time division multiplexing}
 \acro{TDMA}{time division multiple access}
  \acro{TDoA}{time difference of arrival}
 \acro{TEID}{tunnel endpoint identifier}
 \acro{TLS}{transport layer security}
 \acro{TNL}{transport network layer}
  \acro{ToA}{time of arrival}
 \acro{TP}{throughput}
 \acro{TPC}{transmit power control}
 \acro{TPM}{trusted platform module}
 \acro{TR}{transition region}
 \acro{TS}{tabu search}
 \acro{TSG}{technical specification group}
 \acro{TTG}{transmit/receive transition gap}
 \acro{TTI}{transmission time interval}
 \acro{TTT}{time-to-trigger}
 \acro{TU}{typical urban}
 \acro{TV}{television}
 \acro{TWXN}{two-way exchange network}
 \acro{TX}{transmit}
 \acro{UARFCN}{UTRA absolute radio frequency channel number}
 \acro{UAV}{unmanned aerial vehicle}
 \acro{UCI}{uplink control information}
 \acro{UDP}{user datagram protocol}
 \acro{UDN}{ultra-dense network}
 \acro{UE}{user equipment}
 \acro{UGS}{unsolicited grant service}
 \acro{UICC}{universal integrated circuit card}
 \acro{UK}{united kingdom}
 \acro{UL}{uplink}
 \acro{UMA}{unlicensed mobile access}
 \acro{UMi}{urban micro}
 \acro{UMTS}{universal mobile telecommunication system}
 \acro{UN}{United Nations}
 \acro{URLLC}{ultra-reliable low-latency communication}
 \acro{US}{upstream}
 \acro{USIM}{universal subscriber identity module}
 \acro{UTD}{theory of diffraction}
 \acro{UTRA}{UMTS terrestrial radio access}
 \acro{UTRAN}{UMTS terrestrial radio access network}
 \acro{UWB}{ultra wide band}
 \acro{VD}{vertical diffraction}
 \acro{VDFP}{vertical dynamic frequency planning}
 \acro{VDSL}{very-high-bit-rate digital subscriber line}
 \acro{VeNB}{virtual eNB}
 \acro{VeNodeB}{virtual eNodeB}
 \acro{VIC}{victim cell}
 \acro{VLR}{visitor location register}
 \acro{VNF}{virtual network function}
 \acro{VoIP}{voice over IP}
 \acro{VoLTE}{voice over LTE}
 \acro{VPLMN}{visited PLMN}
 \acro{VR}{visibility region}
  \acro{VRAN}{virtualized radio access network}
 \acro{WCDMA}{wideband code division multiple access}
 \acro{WEP}{wired equivalent privacy}
 \acro{WG}{working group}
 \acro{WHO}{world health organisation}
 \acro{Wi-Fi}{Wi-Fi}
 \acro{WiMAX}{wireless interoperability for microwave access}
 \acro{WiSE}{wireless system engineering}
 \acro{WLAN}{wireless local area network}
 \acro{WMAN}{wireless metropolitan area network}
 \acro{WNC}{wireless network coding}
 \acro{WRAN}{wireless regional area network}
 \acro{WSEE}{weighted sum of the energy efficiencies}
 \acro{WPEE}{weighted product of the energy efficiencies}
 \acro{WMEE}{weighted minimum of the energy efficiencies}
 \acro{X2}{x2}
 \acro{X2-AP}{x2 application protocol}
 \acro{ZF}{zero forcing}

 \end{acronym}

\end{document}